\documentclass[aps,prl,pdflatex,twocolumn,notitlepage,superscriptaddress]{revtex4-1}
\usepackage[compat=1.0.0]{tikz-feynman}
\pdfoutput=1

\usepackage{amssymb}
\usepackage{multirow}
\usepackage{bm}
\usepackage{amsmath}
\usepackage{graphicx}
\usepackage{epstopdf}
\usepackage{subfigure}
\usepackage{natbib}
\usepackage{comment}
\usepackage{epsfig}
\usepackage{amsfonts}
\usepackage{mathrsfs}
\usepackage[toc,page,title,titletoc,header]{appendix}
\usepackage[colorlinks,linkcolor=blue,citecolor=blue,anchorcolor=blue]{hyperref}

\usepackage{braket}
\DeclareRobustCommand{\Eq}[1]{Eq.~(\ref{#1})}

\usepackage{dsfont,amsthm,amsbsy}

\def\Tr{{\rm Tr}}

\usepackage{fancyhdr}
\usepackage{xcolor}
\usepackage{ulem}
\usepackage{bbm}
\usepackage{xcolor}

\newcommand{\bk}{\mathbf{k}}
\newcommand{\br}{\mathbf{r}}

\newcommand{\vM}{\vec{M}}
\newcommand{\BZ}{\mathrm{BZ}}

\begin{document}

\title{A quantum geometric mechanism for chiral domain wall metastability: \\
Application to twisted transition-metal dichalcogenides}

\author{Nisarg Chadha}
\affiliation{Department of Physics, Harvard University, Cambridge MA 02138, USA}

\author{Qiang Gao}
\affiliation{Department of Physics, Harvard University, Cambridge MA 02138, USA}

\author{Eslam Khalaf}
\email{eslam_khalaf@fas.harvard.edu}
\affiliation{Department of Physics, Harvard University, Cambridge MA 02138, USA}

\author{Zhaoyu Han}
\email{zhan@fas.harvard.edu}
\affiliation{Department of Physics, Harvard University, Cambridge MA 02138, USA}
\begin{abstract}
Band topology can have an imprint on the excitations of a ferromagnet. 
A known example is quantum Hall ferromagnets and their lattice analogs; when both flavors have the same Chern number $C$, a smooth skyrmion texture binds charge $-eC$ per unit winding. 
Here, we consider instead the case of conjugate Chern bands related by time-reversal.
We show that, despite the vanishing net charge response, a smooth texture can be associated with a dipole response---a domain wall (DW) with an in-plane winding along its length can bind a nonzero dipole density transverse to the wall. 
The strength of this dipole density is controlled by a dimensionless coefficient $c_G$.
Although not quantized, the geometric dipole coefficient $c_G$ is a moment of the second Chern form of the occupied projector in mixed (momentum and order-parameter) space and is generally nonzero. 
The dipole-decorated DW can thus become metastable at a finite radius due to the competition between dipolar repulsion and the usual surface tension, even at a finite Zeeman field.
In a realistic model of twisted MoTe$_2$, we find that $c_G$ drops sharply across a transition within the valley-polarized (VP) phase from a $C=1$ to a $C=0$ ferromagnet. 
This naturally explains recent pump-probe experiments~\cite{exp} at hole filling $\nu=1$, in which a long-lived excitation survives reverse fields far exceeding the saturation field but disappears at an intermediate displacement field despite only weak changes in conventional magnetic diagnostics.
Metastable spin textures thus serve as a sensitive probe of band quantum geometry, and as an intrinsic bottleneck for fast optical control of moiré ferromagnets in Chern-conjugate bands.
\end{abstract}

\maketitle

\textit{Introduction.---}
Moir\'e materials provide a versatile platform for realizing narrow bands whose Bloch wave functions carry nontrivial topology and quantum geometry (QG), usually together with an approximate spin, valley, and/or flavor degeneracy~\cite{PhysRevX.14.041004, PhysRevX.10.031034, PhysRevB.103.205414, doi:10.1126/sciadv.abf5299, PhysRevLett.122.246401, PhysRevB.109.125141, PhysRevB.105.L140506, 2pff-xy6n, An2025SecondBandFerromagnetismMoTe2, Xu2025SecondBandMoTe2, PhysRevLett.132.146401, QiuWu2025, Qiu2025ChiralExcitonicQuantumGeometry, Li2024ElectricallyTunedTopologyMagnetismMoTe2, Li2026ReviewTwistedHomobilayerTMD, PhysRevB.107.075424}.
At integer fillings, exchange interactions can polarize the flavor degree of freedom so that the system becomes a ferromagnet.
A natural question is what imprint the QG of the underlying bands leaves on the resulting magnet, and what sets it apart from a conventional ferromagnet of localized spins or of topologically trivial bands.

The answer is well understood in the case of a ferromagnet formed in bands carrying the \textit{same} Chern number $C$ for both flavors.
Such states are lattice analogs of quantum Hall ferromagnets and inherit their hallmark: charge density is tied to the winding of the order parameter, so that a smooth skyrmion binds a charge $-eC$ per unit winding number~\cite{Sondhi1993}. 
That the charge is carried by a spin texture rather than by individual particles has observable consequences for transport~\cite{Schmeller1995}, thermodynamics~\cite{Barrett1995,Bayot1996,Yu2023}, and the response to a magnetic field~\cite{Melinte1999,Yu2023}.

Far less is understood about ferromagnets realized in conjugate Chern bands, where time reversal enforces opposite Chern numbers for the two flavors, and more generally for any ferromagnet formed in bands with vanishing net topology but nontrivial QG.
The former situation arises for twisted transition-metal dichalcogenide (tTMD) homobilayers such as twisted MoTe$_2$ (tMoTe$_2$) and tWSe$_2$, where spin-orbit coupling locks spin to valley, leaving two flavors with opposite Chern numbers related by time reversal~\cite{Cai2023,Park2023,Zeng2023,Xu2023}.
Here the naive quantum-Hall-ferromagnet logic does not apply: a texture that rotates between the two flavors cannot be assigned a single Chern number, and the 
charge obtained by averaging over the two flavors cancels.
It is therefore not obvious whether QG leaves any imprint at all on the excitations of such a magnet, nor which geometric data would control it.
Answering this requires going beyond the Berry curvature and Fubini--Study metric of a fixed occupied band, since these are evaluated at a single order-parameter orientation, whereas a texture explores the entire family of occupied-band projectors.

The question is especially pressing in tTMDs, where numerical studies find closely competing valley-polarized (VP) ferromagnets with or without nontrivial band topology, and inter-valley coherent (IVC) states~\cite{Devakul2021, Reddy2023, WangPRX2023, WangHigherFilling2023, WangDiverseMagnetic2023, ZhangNatComm2024, WangVafek2024, Wu2020QuantumGeometryFlatbandFerromagnetism, Pan2020TopologyHubbardDMIWSe2, Wu2019TopologicalInsulatorsTMDHomobilayers}.
The small energy splittings between these orders imply a soft and highly tunable order-parameter landscape, in which domain walls (DWs) and other metastable textures can locally interpolate between nearby VP and IVC configurations.

An additional motivation comes from recent resonant pump-probe experiments on tMoTe$_2$ at hole filling $\nu=1$, which reported a striking non-equilibrium excitation~\cite{exp}.
The excitation is long-lived on the scale of tens of microseconds, appears only above a threshold pump fluence, survives reverse magnetic fields several times larger than the saturation field, and disappears sharply around $T\simeq 3.8$~K, far below the magnetic transition temperature $T_c\sim 12$~K~\cite{UniversalMagnetic2025}.
It also exists only in a narrow range of displacement field $D$ inside the ferromagnetic region.
These observations pose a puzzle: the excitation behaves neither like dilute magnons, which should dominate in the low-density limit, nor like ordinary Ising domain walls (DWs), whose stability should track conventional magnetic diagnostics such as coercive field and $T_c$.
This suggests a texture whose stability is set by something other than the usual magnetic anisotropies.

In this Letter, we propose and develop a theory of metastable dipole-decorated chiral DWs in Ising ferromagnetic insulators with non-trivial QG.
A chiral DW is a closed DW across which the in-plane component of the spin/valley order parameter winds by a non-zero integer, which is topologically protected as long as the texture is large compared with microscopic length scales.
Equivalently, it may be viewed as a ring-shaped skyrmion whose topological density is localized on a ring.

Our central result is that such a chiral DW binds a dipole density whose strength is controlled by a dimensionless number we call the geometric dipole coefficient $c_G$---a moment of the mixed real-space--momentum-space second Chern form of the occupied-band projector on the enlarged parameter space $(k_x,k_y,\theta,\phi)$.
Here $(\theta,\phi)$ are the Bloch-sphere angles characterizing the spin/valley polarization direction: $\theta=0,\pi$ denote the two VP states related by time reversal, while $\phi$ parameterizes the spontaneously chosen `in-plane' IVC order on the equator $\theta=\pi/2$.
Time reversal forces the integral of this second Chern form over the order-parameter sphere to vanish, but the wall profile weights it into a nonzero $c_G$.
For an idealized $\bk$-independent flavor texture in conjugate Chern bands with Chern number $\pm C$, we obtain $|c_G| = |C|/4\pi$: the dipole is fixed by the same Chern number that sets the skyrmion charge when both flavors share it.

The resulting dipolar response generates an energy cost that grows as the wall shrinks, thereby stabilizing the texture against collapse even in a reverse Zeeman field.
The mechanism requires only an internal relative $U(1)_{\rm spin}$ symmetry, a $\mathbb{Z}_2^{\rm TR}$ time-reversal symmetry, and an occupied projector that varies over both momentum and order-parameter space; conjugate Chern bands are its cleanest realization, but topologically trivial bands with nontrivial QG can support it as well.

Numerically, we find that $c_G$ is large in the $C=1$ VP regime of tMoTe$_2$ but drops sharply across an internal transition to a $C=0$ VP ferromagnet.
Within the $C=1$ VP ground-state regime we further find a transition between two distinct lowest-energy IVC states, which also produces an abrupt jump in $c_G$.
The dipolar response is thus sensitive both to the Ising ferromagnetic ground state and to the $U(1)$ excited state explored by the texture.
This structure resolves the experimental puzzle above: the long-lived excitation seen in Ref.~\cite{exp} appears only in a narrow window of $D$ and vanishes across an internal transition while the coercive field and $T_c$ remain largely unchanged.

\textit{Textures in flavor ferromagnets.---}
We consider systems with two spin/valley flavors related by time-reversal symmetry, along with a relative $U(1)_{\rm spin}$ conservation.
We use `spin' and `valley' interchangeably throughout: the theory requires nothing beyond this time-reversal relation (which ensures the net charge response vanishes for a smooth texture), whereas in tMoTe$_2$ and similar TMD systems the two flavors are additionally spin-valley locked.
Time reversal maps the bands of one flavor onto those of the other, forcing their Chern numbers to be opposite and the net topology to vanish.

At integer fillings, exchange interactions commonly favor flavor ferromagnetism, with the orientation of order selected by the competition between band dispersion, quantum geometry, and interaction anisotropies.
Numerical calculations on tTMDs find the VP and IVC orders to be nearly degenerate~\cite{WangDiverseMagnetic2023,WangPRX2023}.
The former is an easy-axis ferromagnet; the latter is an easy-plane ferromagnet in which the two flavors hybridize, breaking the internal $U(1)_{\rm spin}$ symmetry.

In the regime where these orders compete closely, it is useful to embed them in an enlarged order parameter $\vM\in S^2$, where $M_z$ distinguishes the two Ising VP states, and $(M_x,M_y)$ captures the amplitude and phase of the IVC order.
Embedding nearly degenerate flavor orders in a common order-parameter manifold is standard in moiré magnets~\cite{PhysRevX.10.031034,Khalaf2020SoftModes,Kwan2022Skyrmions}, and parallels the treatment of competing $\nu=0$ orders in graphene quantum Hall ferromagnets~\cite{Kharitonov2012}.
We shall employ such a construction to describe the low-energy textures interpolating between the VP and IVC saddle points.

A slowly varying texture $\vM(\br)$ is described by a family of occupied-band projectors $P[\bk;\vM(\br)]$.
We construct this family from two reference Slater determinants obtained by symmetry-constrained Hartree--Fock (HF) for $\hat H=\hat H_0+\hat V$, where the quadratic $\hat H_0$ defines the active Bloch bands and $\hat V$ contains repulsive density--density interactions.
HF within the $U(1)_{\rm spin}$-symmetric sector gives the VP polarized state $\rho_{+}^{\rm VP}$, with $\rho_{-}^{\rm VP}$ obtained by time-reversal. 
The IVC states are obtained by imposing the non-Kramers time-reversal symmetry
$\mathcal{T}' = \tau_x K$, which combines the Kramers time reversal
$\mathcal{T}=i\tau_y K$ with a $\pi$ rotation in flavor space (here
$\tau_{x,y,z}$ are Pauli matrices in flavor space and $K$ is complex
conjugation), and satisfies $\mathcal{T}'^2=+1$.
Applying $U(1)_{\rm spin}$ rotations to a reference state $\rho^{\rm IVC}_0$ generates the symmetry-related family $\rho^{\rm IVC}_{\phi}$ with $\phi\in[0,2\pi)$.
We use these converged HF states to define the corresponding self-consistent HF Hamiltonians $\hat{H}_{\rm HF}[\rho_{\pm}^{\rm VP}]$ and $\hat{H}_{\rm HF}[\rho_{\phi}^{\rm IVC}]$.
Here, $\hat{H}_{\rm HF}[\rho]$ is the quadratic Hamiltonian obtained by the HF decoupling of the quartic interacting Hamiltonian against $\rho$.

We use these HF Hamiltonians to construct a family of trial Hamiltonians parametrized by the $\vM$ orientation $(\theta, \phi)\in[0,\pi]\times[0,2\pi)$:
\begin{equation}
\label{eq:interpolation}
\hat H_{\rm MF}(\theta,\phi)
=
\hat H_0+
\left|\cos\theta\,\right|\delta \hat H^{\rm VP}_{\mathrm{sign}(\cos\theta)}
+
\sin\theta\,\delta \hat H^{\rm IVC}(\phi), 
\end{equation}
where:
\begin{equation}
\label{eq:interpolation2}
\begin{aligned}
\delta \hat H^{\rm VP}_{\pm} &= \hat H_{\rm HF}[\rho^{\rm VP}_{\pm}]-\hat H_0,\\
\delta \hat H^{\rm IVC}(\phi) &= \hat H_{\rm HF}[\rho_\phi^{\rm IVC}]-\hat H_0.
\end{aligned}
\end{equation}

For each $(\theta,\phi)$ we numerically diagonalize the quadratic Hamiltonian
$\hat{H}_{\rm MF}(\theta,\phi)$ to obtain the variational projector at that
point on $S^2$.
The poles $\theta=0,\pi$ denote the two oppositely polarized VP states, while
$\theta=\pi/2$ is the IVC equator, on which $\phi$ remains free.
We use $|\cos\theta|$ and $\sin\theta$ rather than $\cos^2\theta$ and $\sin^2\theta$ to preclude fine-tuned zeros of the geometric response that would arise from the parametrization rather than from the underlying symmetries: this choice keeps $\partial_{\theta}\hat{H}_{\rm MF}$ generically nonvanishing on $[0,\pi]$, while the absolute value keeps the weight positive so that $\theta=0,\pi$ reproduce the two VP states exactly.

\begin{figure}[t]
\includegraphics[width=\columnwidth]{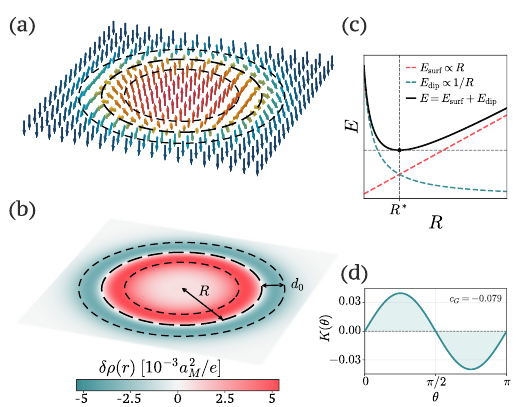}
\centering
\caption{
\textbf{Schematic of a quantum-geometric chiral domain wall.}
(a) shows the schematic of a chiral domain wall of the form described by Eq.~\eqref{eq:chiralDW} with an in-plane winding $N_w = 1$ in an Ising ferromagnet.
(b) shows the induced charge density for the projector given by Eq.~\eqref{eq:projector} with the chiral domain wall in Eq.~\eqref{eq:chiralDW} with $R=14\ell_B$ and $d_0 = 4\ell_B$ for a system of conjugate lowest Landau levels.
(c) shows the competition in the wall energy arising from the surface tension $2\pi\sigma R$ and dipolar repulsion $\sim (N_wc_G)^2/R$, giving a finite-radius metastable minimum.
(d) $K(\theta)$ vs $\theta$ for the spin texture described in Eq.~\eqref{eq:projector} for a setup of two conjugate lowest Landau levels with a mesh size of $N_k = 18$. The calculated $c_G$ agrees with the ideal value $-1/4\pi$ evaluated from Eq.~\eqref{eq:chargeideal}.
}
\label{fig:mechanism}
\end{figure}

\textit{Dipole-decorated chiral domain walls.---}
An easy-axis ferromagnet supports domain walls with characteristic width $d_0$ and line tension $\sigma$.
Across the wall, $M_z$ changes sign and an in-plane component develops in its core.
For a closed wall, its phase may wind along the arclength $\ell$,
\begin{align}
N_w
=
\frac{1}{2\pi}
\oint d\ell\,\partial_\ell\phi
\in\mathbb Z.
\end{align}
Domain walls with $N_w\neq0$ are chiral, the sign of $N_w$ specifying the handedness (Fig.~\ref{fig:mechanism}a shows a schematic with $N_w=1$).

For a circular wall centered at radius $R$, we use the following ansatz in the polar coordinate $(r,\alpha)$:
\begin{align}
\vM(r,\alpha)
=
\left(
\sin\theta(r)\cos N_w\alpha,
\sin\theta(r)\sin N_w\alpha,
\cos\theta(r)
\right),
\label{eq:chiralDW}
\end{align}
with the following profile ($d_0$ is the domain wall width)
\begin{align}
\theta(r)
=
2\tan^{-1}\exp\left(\frac{2(r-R)}{d_0}\right).
\end{align}
Our texture resembles a skyrmion whose topological density is localized on a ring. However, unlike Chern ferromagnets, where such a ring would bind a net charge, we intuitively expect equal and opposite charge on the inside and outside of the ring since the opposite spin bands experience opposite Chern numbers. Instead, we expect a net dipole moment localized at the domain wall, as we will show below.

The leading geometric charge response is given by the mixed real-space--momentum-space second-Chern form~\cite{QiHughesZhang2008,XiaoShiNiu2009,ZhaoGaoXiao2021},
\begin{align}
\delta\rho(\br)
=
-\frac{e}{8\pi^2}
\int_{\BZ}d^2k\,
\epsilon^{abcd}
\Tr\left[
P(\partial_aP)(\partial_bP)(\partial_cP)(\partial_dP)
\right],
\label{eq:second_chern}
\end{align}
where $a,b,c,d\in\{k_x,k_y,x,y\}$ and
$P=P[\bk;\vM(\br)]$.
This response depends only on the occupied subspace but not on the corresponding band energies, and is thus geometric.

For the ansatz in Eq.~\eqref{eq:chiralDW},
\begin{align}
\delta\rho(r,\alpha)
=
-eN_w\frac{\partial_r\theta}{r}
K\!\left(\theta(r),\phi(\alpha)\right),
\label{eq:rho_kernel}
\end{align}
where
\begin{align}
K(\theta,\phi)
=
\frac{1}{8\pi^2}
\int_{\BZ}d^2k\,
\epsilon^{abcd}
\Tr\left[
P(\partial_aP)(\partial_bP)(\partial_cP)(\partial_dP)
\right],
\label{eq:kernel}
\end{align}
with $a,b,c,d\in\{k_x,k_y,\theta,\phi\}$.
Equation~\eqref{eq:kernel} is the second Chern density of the occupied-band projector on the four-dimensional parameter space $(k_x,k_y,\theta,\phi)$, with the momentum directions integrated out, so that $K$ is a density on the
order-parameter sphere.
Because $\epsilon^{abcd}$ antisymmetrizes over all four directions, every term
carries exactly one derivative along each of $k_x$, $k_y$, $\theta$ and $\phi$:
$K$ probes only the mixed momentum--order-parameter components of the
non-Abelian curvature, and vanishes whenever the occupied states are momentum independent, as is the case for a ferromagnet composed of localized moments. 
Strictly speaking, Eq.~\eqref{eq:second_chern} assumes a finite spectral gap throughout the interpolation, whereas this condition must fail for certain $\vM$ when connecting VP states with opposite Chern numbers. 
These gap closings correspond to the chiral edge modes localized along the domain wall, near which higher-order and nonadiabatic corrections can modify the detailed charge redistribution.

Flavor $U(1)$ and time-reversal symmetry constrain the kernel\footnote{Both properties follow from the symmetry action on the projector
family: a $U(1)_{\rm spin}$ rotation conjugates $P$ by a $\bk$- and
$\theta$-independent unitary, which cancels cyclically in the trace, while
time reversal sends $(\bk,\theta,\phi)\to(-\bk,\pi-\theta,\phi+\pi)$, under
which only the single $\partial_\theta$ appearing in each term of
Eq.~\eqref{eq:kernel} changes sign.}:
$K$ is independent of $\phi$, so that $K(\theta,\phi)\equiv K(\theta)$, and
\begin{align}
K(\theta) = -K(\pi-\theta).
\end{align}

The induced charge therefore changes sign across the wall leading to a vanishing net charge, as shown in Fig.~\ref{fig:mechanism}b for the case of conjugate lowest Landau levels.
Its leading multipole is a radial dipole moment per unit length,
\begin{align}
p_0
=
\int_0^\infty dr\,(r-R)\delta\rho(r)
\simeq
-e\frac{N_wd_0}{R}c_G,
\label{eq:dipole}
\end{align}
where the final expression holds for $R\gg d_0$, and
\begin{align}
c_G
=
\int_0^\pi d\theta\,
K(\theta)
\ln\tan\left(\frac{\theta}{2}\right).
\label{eq:cG}
\end{align}
The coefficient $c_G$ is the central quantum-geometric quantity of this work.
It depends on the full projector family over
$(k_x,k_y,\theta,\phi)$, rather than only on the Berry curvature or Fubini--Study metric at a fixed order-parameter orientation.

\textit{Metastability.---}The dipole line density \Eq{eq:dipole} produces a Coulomb energy scaling as $N_w^2/R$, which prevents the domain from shrinking.
Including the wall tension $\sigma$ and a background magnetic field bias $B_Z$ gives~\footnote{A caveat here is that the in-plane stiffness term generically also contributes to an energy scaling as $N_w^2/R$. However, its value is rather small in tMoTe$_2$~\cite{QiuWu2025}. }
\begin{equation}
E(R)
=
\frac{2\pi e^2(N_wc_G)^2}{\epsilon R}
+
2\pi\sigma R
+
2\pi\mu B_ZR^2,
\label{eq:energy}
\end{equation}
where $\epsilon$ is the dielectric constant and $\mu$ is the magnitude of the magnetization-density of a single VP state\footnote{$\mu$ receives both spin and orbital contributions.
Spin--valley locking fixes the spin part, while the orbital part arises from the valley magnetic moment of the Chern band; for tTMDs we compute it to be
$\approx 1.5\,\mu_B$ per hole.}. 
$B_Z>0$ penalizes the polarization inside the domain.
At zero field,
\begin{align}
\label{eq:Rstar}
    R_{N_w}^\ast
&=
\frac{|eN_wc_G|}{\sqrt{\epsilon\sigma}},
&
E_{N_w}^\ast
&=
4\pi|eN_wc_G|\sqrt{\frac{\sigma}{\epsilon}},
\end{align}
as shown in Fig.~\ref{fig:mechanism}c. Although the energy grows with $|N_w|$, winding is topologically protected for large enough domains and cannot jump. 
Note that the optimal radius grows linearly with $|N_w|$, suggesting metastability against shrinking. 
Thus, our metastability mechanism has two ingredients: the topological winding cannot be undone for large enough domain walls whose metastability against shrinking is ensured by $c_G$ and the generated dipole moment.

The above analysis has two important qualifications for its validity. First, the domain wall width $d_0$ and the optimal tangential winding wavelength $\lambda \equiv 2\pi R_{N_w}^*/|N_w| = \frac{2\pi|e\,c_G|}{\sqrt{\epsilon\sigma}}$ must both be larger than characteristic microscopic length scale $\ell_\text{UV}$ (e.g. moir\'e lattice spacing) in order for the continuum description to be valid. Second, the winding $|N_w|$ should be sufficiently large, such that the wall can reach a radially resolved regime where $R^*$ is large compared to $d_0$. Both conditions are to make sure that the topological structure in the texture is mesoscopically stable.

For a fixed $N_w$, domain wall collapse requires the domain to shrink to $R\sim d_0$, giving rise to a barrier. Conversely, nucleation requires energy to overcome the barrier, implying a finite critical pump fluence to create such metastable domains. A reverse magnetic field opposing the polarization inside the domain shifts the optimal $R^*$ to a smaller $R$, but the domain remains metastable as long as the field-dependent minimum satisfies $R^\ast\gtrsim d_0$. This explains why a chiral wall can survive fields far larger than the saturation field of ordinary domains.

\textit{An ideal limit for $c_G$.---} 
It is instructive to consider an idealized two-band limit with a Chern band $\ket{u_s(\bk)}$ of Chern number $C_s$ from each flavor $s=\pm$, and take a simplified ansatz where the
texture acts as a momentum-independent rotation within the flavor space they span:
\begin{align}
P(\bk;\br)
&=
\sum_{ss'}
\ket{u_s(\bk)}
p_{ss'}(\br)
\bra{u_{s'}(\bk)},
\nonumber\\
p(\br)
&=
\frac{1}{2}
\left[
1+\vM(\br)\cdot\vec\tau
\right].
\label{eq:projector}
\end{align}
Equation~\eqref{eq:second_chern} then reduces to
\begin{equation}
\begin{aligned}
\delta\rho_{\rm ideal}(\br)
&=
-e\,\bar C(\br)\rho_{\rm sk}(\br),
\label{eq:adiabatic}\\
\rho_{\rm sk}(\br)
&=
\frac{1}{4\pi}
\vM\cdot
\left(
\partial_x\vM\times\partial_y\vM
\right),
\\
\bar C(\br)
&=
C_+\cos^2\frac{\theta}{2}
+
C_-\sin^2\frac{\theta}{2}.
\end{aligned}
\end{equation}

For a conventional quantum Hall ferromagnet with $C_+=C_-=C$, this gives the familiar result
$\delta\rho=-eC\rho_{\rm sk}$~\cite{Sondhi1993}.
For time-reversal-conjugate bands, $C_+=-C_-=C$, one instead obtains
\begin{align}
\delta\rho_{\rm ideal}(\br)
=
-eC\cos\theta(\br)\rho_{\rm sk}(\br).
\end{align}
The charge density therefore changes sign across the wall, producing a dipole rather than a net charge.
For Eq.~\eqref{eq:chiralDW},
\begin{align}
\delta\rho_{\rm ideal}(r)
=
-\frac{eCN_w}{4\pi r}
\cos\theta(r)\sin\theta(r)\partial_r\theta,
\label{eq:chargeideal}
\end{align}
which corresponds to
\begin{align}
K(\theta)
=
\frac{C}{8\pi}\sin2\theta,
\qquad
c_G
=
-\frac{C}{4\pi},
\end{align}
as shown in Fig.~\ref{fig:mechanism}d.

Strictly speaking, bands with different Chern numbers cannot be connected by a globally smooth, momentum-independent hybridization~\cite{xie2024theory}.
The factorized projector in Eq.~\eqref{eq:projector} should therefore be regarded as a rudimentary, analytically transparent limit rather than a globally well-defined microscopic construction.
Nevertheless, our Hartree--Fock calculations for conjugate lowest Landau levels and Aharonov--Casher bands~\cite{Aharonov1979GroundStateSpinHalf,ShiKhalafMacDonald2024} yield $c_G$ close to $-1/(4\pi)$ (within about $30\%$ relative error) over a broad parameter range, demonstrating the usefulness of this ideal estimate (see Supplemental material).

\textit{Chiral domain walls in tMoTe$_2$.---}We model tMoTe$_2$ using the conventional two-layer moir\'e continuum model introduced in Ref.~\cite{Wu2019TopologicalInsulatorsTMDHomobilayers}.
Strong spin-orbit coupling in TMDs locks the spin and valley of the low-energy bands, giving rise to two time-reversal-related flavors.
The moir\'e valence bands arise from layer-dependent periodic potentials and spatially modulated interlayer tunneling, which are constrained by the lattice and internal symmetries of the model~\cite{Wu2019TopologicalInsulatorsTMDHomobilayers}.
We use the parameters of Ref.~\cite{WangDiverseMagnetic2023}, obtained by fitting the continuum model to first-principles calculations, and tune the twist angle $\theta_M$ and the vertical displacement field $u_D$.
Interactions are incorporated through the standard projection of a dual-gate screened Coulomb interaction onto the two highest valence bands per flavor.
The continuum Hamiltonian and interaction projection follow standard constructions~\cite{Wu2019TopologicalInsulatorsTMDHomobilayers, WangDiverseMagnetic2023, WangHigherFilling2023}.
Their explicit forms, numerical parameters, and symmetry conventions are detailed in the End Matter.

\begin{figure}[t]
\centering
\includegraphics[width=\columnwidth]{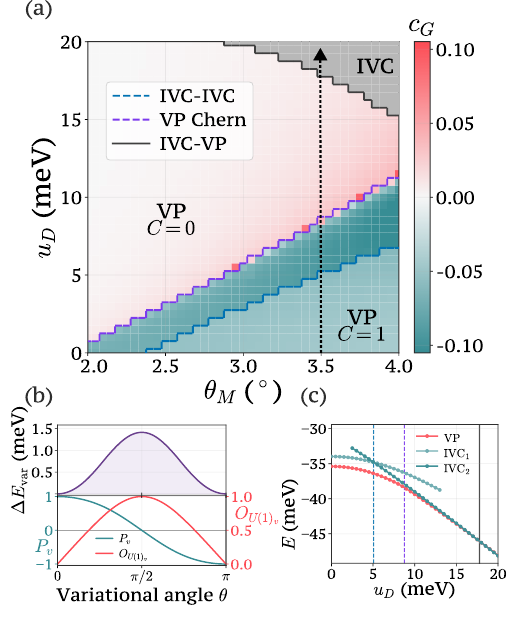}
\caption{
\textbf{Hartree-Fock calculations for $c_G$ in the continuum model for tMoTe$_2$.}
(a) shows the values of $c_G$ extrapolated from HF calculations as a function of varying twist angle $\theta_M$ and displacement field $u_D$. 
(b) shows the energy expectation value per moir\'e unit cell and the VP ($P_v$) and IVC ($O_{U(1)_v}$) order parameters for the variational state parametrized by $\theta$ and given by the ground state of the trial Hamiltonian $\hat{H}_{\rm MF}(\theta,\phi)$ given in Eq.~\eqref{eq:interpolation} for the tMoTe$_2$ continuum model with twist angle $\theta_M = 3.5^{\circ}$ and $u_D = 0$. The variational energy is given with respect to the VP state.
(c) HF energy per unit cell for the different phases along a line cut shown by the arrow in (a).
}
\label{fig:tMoTe2}
\end{figure}

Comparing self-consistent HF solutions at hole filling $\nu=1$ across the
$(\theta_M,u_D)$ plane reveals a close competition between the
interaction-driven VP and IVC states, with a VP--IVC phase boundary spanning
experimentally tunable parameters, as shown in Fig.~\ref{fig:tMoTe2}a.
Within the VP region, we also find a topological transition at which the Chern
number of the occupied hole HF band changes from $\pm1$ to $0$.
The energetic proximity of the VP and IVC orders motivates
Eq.~\eqref{eq:interpolation} as a variational description of spatially varying
textures that locally interpolate between them.

At each $(\theta_M, u_D)$, we obtain the VP and IVC reference states $\rho^{\rm VP}_{\pm}$ and $\rho^{\rm IVC}_{\phi}$ respectively, entering Eq.~\eqref{eq:interpolation} from symmetry-constrained HF calculations.
Throughout the entire parameter range considered in Fig.~\ref{fig:tMoTe2}a, the energy difference between the lowest-energy IVC and VP states stays less than $2\,\mathrm{meV}$ per unit cell, which is significantly smaller than the naive anisotropy scale $\gtrsim 6.4\,$meV inferred from the magnon gap~\cite{QiuWu2025}.

Our main numerical result is shown in Fig.~\ref{fig:tMoTe2}a. 
At small displacement field, the VP state is a $C=1$ Chern ferromagnet with a large $c_G$.
At intermediate displacement field, the system remains VP but undergoes a topological transition to a topologically trivial $C=0$ ferromagnet, across which $c_G$ changes sign and drops significantly in magnitude. 
At even larger displacement fields, the Hartree-Fock ground state becomes IVC, and DWs cease to be the relevant excitations.
Thus, $c_G$ diagnoses qualitative changes within the VP ferromagnet that are largely invisible to conventional magnetic probes such as the Curie temperature, coercive field, and static magneto-optical hysteresis, which largely remain invariant across the intra-VP transition.

Within the region where the HF ground state is a $|C|=1$ VP state, the lowest-energy solution in the $\mathcal T'$-symmetric sector switches discontinuously between two metastable IVC branches, as shown in Fig.~\ref{fig:tMoTe2}c.
Although both occupied IVC HF bands are Chern trivial, their order parameters transform differently under threefold rotation and their Wannier charge centers lie at inequivalent positions within the moir\'e unit cell.
The low-$u_D$ branch is centered at the MM stacking position, whereas the high-$u_D$ branch sits at the MX/XM position and therefore realizes an obstructed atomic insulator~\cite{PhysRevB.86.115112,po2017symmetry}.
Here, M and X denote the transition-metal (Mo) and chalcogen (Te) sublattices, respectively, while MX and XM distinguish which layer supplies each sublattice~\cite{Wu2019TopologicalInsulatorsTMDHomobilayers}.
This distinction is also reflected in their momentum-space IVC textures.
The transition persists for all meshes studied, $N_k=18,\ldots,24$, and is accompanied by a discontinuous jump in $c_G$.
A more detailed characterization of the two IVC phases will be presented elsewhere.

Our results offer a natural explanation of the recent experiment~\cite{exp}, where the conventional magnetic properties change little as the displacement field crosses the boundary at which the long-lived excitation disappears.
The same tuning drives the VP state from $C=1$ with large $c_G$ to $C=0$ with much smaller $c_G$.
Because metastability requires $\lambda=\frac{2\pi|e\,c_G|}{\sqrt{\epsilon\sigma}}\gtrsim\ell_{\rm UV}$, chiral domain
walls are lost as $c_G$ falls while the Ising ferromagnetism is untouched,
making the excitation lifetime a dynamical probe of band quantum geometry where
static magnetic probes are blind.
The two IVC branches imply a further signature: displacement-field sweeps
should produce hysteresis in the domain-wall lifetime if the wall switches
between branches of different $c_G$. 

\textit{Discussion.---}
The main takeaway from our work is that there is an imprint of quantum geometry on the properties of excitations in ferromagnets, even in the absence of net topology, and that such imprint can be encoded in more complicated quantum geometric quantities that mix real and momentum space, thus going beyond conventional measures defined merely in momentum space.
A texture explores the entire family of occupied-band projectors, and the leading charge response it generates is controlled by the second Chern form on the combined momentum--order-parameter manifold. 

Our mechanism also bears on optical switching of moir\'e Chern ferromagnets, where circularly polarized light has recently been used to control integer and fractional Chern states in tMoTe$_2$~\cite{CaiOpticalSwitching2025,OpticalControlTopology2025}.
Chiral DWs enhance the retention of optically written reversed domains, but their nucleation barrier sets a finite writing threshold and their long lifetime slows erasure, which may produce hysteretic or stochastic switching near the
stability boundary.
Both features---long-lived retention and a finite writing threshold---are already present in the excitation reported in Ref.~\cite{exp}.
Quantum geometry thus controls a tradeoff between robust optical memory and rapid reset. We close by noting that our predicted dipole-decorated chiral domain wall should be readily accessible to local probes of magnetization and charge, providing a smoking gun signature of our mechanism. Extending our construction to fractional Chern ferromagnets---where the bound charge should itself
fractionalize---is a natural next step.

{\bf Acknowledgement. }
We thank Chenhao Jin, Richen Xiong, Chenxin Qin for collaborations on the experiment companion paper~\cite{exp}, and Taige Wang, Julian May-Mann and Jonah Herzog-Arbeitman for helpful discussions.
 E.~K. is supported by NSF MRSEC DMR-2308817 through the Center for Dynamics and Control of Materials. Z.~H. is supported by the Simons Investigator award, the
Simons Collaboration on Ultra-Quantum Matter, which
is a grant from the Simons Foundation (651440, Ashvin Vishwanath), and Gordon and Betty Moore Foundation
EPiQS Award 8683.

\bibliographystyle{apsrev4-1}
\bibliography{ref}
\clearpage
\vspace{10em}

\onecolumngrid
\begin{center}
    {\large \textbf{End Matter}}
\end{center}

\renewcommand{\theequation}{A\arabic{equation}}
\setcounter{equation}{0}

\twocolumngrid

\textit{Continuum model for twisted TMD homobilayers.---}
We model the valence bands of twisted TMD homobilayers using the continuum Hamiltonian introduced in Ref.~\cite{Wu2019TopologicalInsulatorsTMDHomobilayers}, with material parameters obtained from the first-principles calculations of Ref.~\cite{WangDiverseMagnetic2023}.
Strong spin-orbit coupling locks spin to valley, leaving two flavors $\tau=\pm$ related by time reversal. 
In the two-layer basis and a $C_3$-symmetric gauge, the single-particle Hamiltonian for flavor $\tau$ is
\begin{equation}
\label{eq:continuum_model}
\begin{aligned}
&\hat{\mathcal H}^{\tau}_0
=
\begin{pmatrix}
-\dfrac{\hbar^2\hat{\bk}^2}{2m^*}
-u_D+\Delta_b(\hat{\br})
&
\Delta_{T,\tau}(\hat{\br})
\\[5pt]
\Delta_{T,\tau}^{\dagger}(\hat{\br})
&
-\dfrac{\hbar^2\hat{\bk}^2}{2m^*}
+u_D+\Delta_t(\hat{\br})
\end{pmatrix},
\\
&\Delta_{b/t}(\br)
=
2V\sum_{j=1,3,5}
\cos\!\left(\mathbf{g}_j\cdot\br\pm\psi\right),
\\
&\Delta_{T,\tau}(\br)
=
w\sum_{j=0}^{2}e^{-i\tau\mathbf{q}_j\cdot\br},
\end{aligned}
\end{equation}
where $\hbar\hat{\bk}=-i\nabla$. Here, $b$ and $t$ label the bottom and top
layers, and the upper (lower) sign in $\Delta_{b/t}$ corresponds to the
bottom (top) layer. The parameters $V$ and $\psi$ determine the
layer-dependent moir\'e potential, $w$ is the interlayer tunneling amplitude,
and $m^*$ is the valence-band effective mass. The layer bias $u_D$
corresponds to an interlayer potential difference $2u_D$ and models a
vertical displacement field.

We choose
\begin{align}
\label{eq:rec_vec}
\mathbf{g}_j
&=
\frac{4\pi}{\sqrt{3}a_M}
\left(
\cos\frac{(j-1)\pi}{3},
\sin\frac{(j-1)\pi}{3}
\right),
\\
\mathbf{q}_j
&=
\frac{4\pi}{3a_M}
\left(
\sin\frac{2\pi j}{3},
-\cos\frac{2\pi j}{3}
\right),
\end{align}
where $\mathbf{g}_j$ are the first-shell moir\'e reciprocal vectors and $\mathbf{q}_j$
are the three moir\'e momentum-transfer vectors entering the interlayer
tunneling. The moir\'e lattice spacing is
\begin{align}
a_M=\frac{a_0}{2\sin(\theta_M/2)}
\simeq\frac{a_0}{\theta_M},
\end{align}
where the final expression applies at small twist angle with $\theta_M$
expressed in radians.

The two flavor Hamiltonians obey
\begin{align}
\mathcal H_0^{-\tau}(\bk)
=
\left[\mathcal H_0^\tau(-\bk)\right]^*.
\label{eq:tprime_continuum}
\end{align}
Within the two-flavor orbital problem, this relation is represented by the
non-Kramers anti-unitary symmetry
$\mathcal T'=\tau_xK$, where $\tau_x$ exchanges the flavors and $K$ denotes
complex conjugation. Unlike physical time reversal for spinful electrons,
$(\mathcal T')^2=+1$; it may be viewed as physical time reversal composed
with a spin rotation. We use Eq.~\eqref{eq:tprime_continuum} to generate the
$\tau=-$ Bloch states from the $\tau=+$ states. This fixes their relative
gauge, which is required when constructing intervalley-coherent form factors
and imposing the $\mathcal T'$ constraint on the IVC HF reference state.

\begin{figure}[t]
\includegraphics[width=\columnwidth]{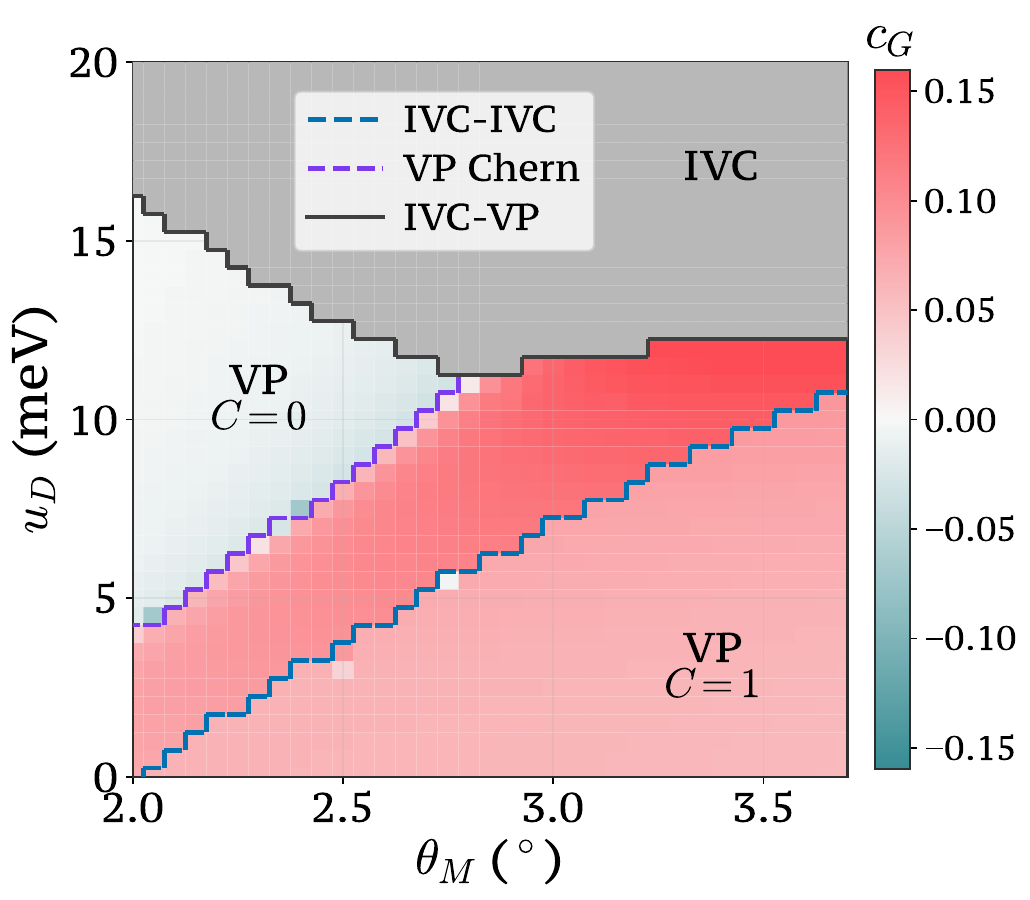}
\caption{
\textbf{Hartree-Fock phase diagram and $c_G$ in twisted WSe$_2$.} This figure shows the HF phase diagram $c_G$ and the computed values of $c_G$ across a range of displacement field $u_D$ and twist angle $\theta_M$. $c_G$ carries the opposite sign compared to tMoTe$_2$ because at these parameters the signs of the Chern numbers in each valley are flipped compared to tMoTe$_2$.
 }
\label{fig:tWSe2}
\end{figure}

We use the continuum-model parameters obtained by fitting first-principles
valence-band calculations in Ref.~\cite{WangDiverseMagnetic2023}:
\begin{equation}
\begin{array}{c|ccccc}
 & a_0\,({\rm \AA}) & m^*/m_e & V\,({\rm meV})
 & \psi & w\,({\rm meV})\\
\hline
{\rm MoTe}_2 & 3.47 & 0.62 & 11.2 & 91^\circ & -13.3\\
{\rm WSe}_2  & 3.32 & 0.43 & 9.0  & -128^\circ & 18.0
\end{array}
\label{eq:ttmd_parameters}
\end{equation}
and vary the twist angle $\theta_M$ and layer bias $u_D$.
For WSe$_2$, our reciprocal-vector convention gives $\psi=-128^\circ$;
this is equivalent to the $\psi=+128^\circ$ convention quoted in
Ref.~\cite{WangDiverseMagnetic2023} after reversing the reciprocal-space
orientation.
Fig. \ref{fig:tWSe2} shows the phase diagram and variation of $c_G$ across a range of $\theta_M$ and $u_D$.

\textit{Projecting interactions.---} We diagonalize Eq.~\eqref{eq:continuum_model} in a truncated plane-wave basis and retain the two highest valence bands from each flavor. 
The stability to the plane-wave cutoff is discussed in the Supplemental Material.
Denoting the corresponding electron operators by $c_{m\tau,\bk}$, with $m=1,2$, the interaction projected into this active subspace is
\begin{align}
\hat H_{\rm int}
&=
\frac{1}{2A}\sum_{\mathbf{q}}V(q):
\bar\rho_e(\mathbf{q})\bar\rho_e(-\mathbf{q}):,
\\
\bar\rho_e(\mathbf{q})
&=
\sum_{\bk,\tau,m,n}
\Lambda^\tau_{mn}(\bk,\mathbf{q})
c^\dagger_{m\tau,[\bk+\mathbf{q}]}c_{n\tau,\bk},
\label{eq:projected_interaction}
\end{align}
where $A$ is the total area, $[\bk+\mathbf{q}]$ denotes folding into the moir\'e
Brillouin zone, and normal ordering is defined relative to charge neutrality.
Writing
$\mathbf G_{\bk,\mathbf{q}}=\bk+\mathbf{q}-[\bk+\mathbf{q}]$, the density form factor is
\begin{align}
\Lambda^\tau_{mn}(\bk,\mathbf{q})
=
\left\langle
u_{m\tau,[\bk+\mathbf{q}]}
\middle|
e^{i\mathbf G_{\bk,\mathbf{q}}\cdot\br}
\middle|
u_{n\tau,\bk}
\right\rangle .
\end{align}
Thus, the projection retains the momentum- and band-dependent Bloch
wavefunction overlaps, including the reciprocal-lattice component of each
momentum transfer.

We use the dual-gate screened Coulomb interaction
\begin{align}
V(q)=
\frac{2\pi e^2}{\epsilon q}
\tanh(qd)\,
e^{-q^2\ell^2/2},
\end{align}
with $\epsilon=16.7$, $d=30\,{\rm nm}$, and the short-distance cutoff $\ell=a_0$ is introduced following Ref.~\cite{WangDiverseMagnetic2023} to account for the diffused nature of the atomic orbitals. 
The uniform Hartree contribution is removed by the neutralizing background. 
At hole filling $\nu=1$, this projected Hamiltonian is used in the symmetry-constrained HF calculations defining the VP and IVC reference states.

\clearpage
\pagestyle{plain}
\onecolumngrid
\begin{center}
{\large \textbf{A quantum geometric mechanism for chiral domain wall metastability: \\
Application to twisted transition-metal dichalcogenides: Supplemental Material}}\\[0.5em]
Nisarg Chadha$^1$, Qiang Gao$^1$, Eslam Khalaf$^1$, and Zhaoyu Han$^1$\\[0.5em]

$^{1}$Department of Physics, Harvard University, Cambridge, Massachusetts 02138, USA\\
\end{center}

\setcounter{section}{0}
\setcounter{equation}{0}
\setcounter{figure}{0}
\setcounter{table}{0}

\renewcommand{\thesection}{S\arabic{section}}
\renewcommand{\theequation}{S\arabic{equation}}
\renewcommand{\thefigure}{S\arabic{figure}}
\renewcommand{\thetable}{S\arabic{table}}

\section{Domain-wall energetics}

For completeness, we first recall the conventional energetics of an easy-axis
ferromagnet,
\begin{align}
E_0[\vM]=\int d^2r
\left[
-a_z M_z^2
+\rho_z(\nabla M_z)^2
+\rho_\perp(\nabla \vM_\perp)^2
+\cdots
\right].
\end{align}
For $\rho_\perp\ll \rho_z$, the domain-wall width and surface tension are
controlled primarily by $a_z$ and $\rho_z$,
\begin{align}
d_0\sim \sqrt{\rho_z/a_z},
\qquad
\sigma\sim \sqrt{a_z\rho_z}.
\end{align}
The in-plane stiffness $\rho_\perp$ gives subleading corrections to the wall
tension and to the energetics of the winding along the wall.

Next we evaluate the dipole interaction energy, taking the dipole moment
density to be localized on the wall over a scale set by the wall width $d_0$.
For a radial dipole line density
\begin{align}
\mathbf{p}(\br)=p_0\hat{\mathbf r}\,\delta(r-R),
\end{align}
the electrostatic energy follows from the bound charge
$\rho_b=-\nabla\!\cdot\!\mathbf p$ after integrating by parts twice,
\begin{align}
E_{\rm dip}
=
-\frac{1}{2}
\int d^2r\,d^2r'\,
p_i(\br)p_j(\br')
\partial_i\partial_j V(|\br-\br'|),
\end{align}
and the interaction between two dipoles separated by $\boldsymbol{\rho}$ is
\begin{align}
U_{12}
=
-\left[
(\mathbf p_1\cdot \mathbf p_2)\frac{V'(\rho)}{\rho}
+
(\mathbf p_1\cdot \hat{\boldsymbol{\rho}})
(\mathbf p_2\cdot \hat{\boldsymbol{\rho}})
\left(
V''(\rho)-\frac{V'(\rho)}{\rho}
\right)
\right]
\equiv
p_0^2\,\mathcal V(\alpha),
\end{align}
where the last equality specializes to two radial dipoles of magnitude $p_0$
on the ring at relative angle $\alpha$, for which
\begin{align}
\rho=2R\sin\frac{\alpha}{2},
\qquad
\mathbf p_1\!\cdot\!\mathbf p_2 = p_0^2\cos\alpha,
\qquad
(\mathbf p_1\!\cdot\!\hat{\boldsymbol{\rho}})
(\mathbf p_2\!\cdot\!\hat{\boldsymbol{\rho}})
=-p_0^2\sin^2\frac{\alpha}{2}.
\end{align}
We use the softened Coulomb kernel
$$V(\rho)=1/\bigl(\epsilon\sqrt{\rho^2+d_0^2}\bigr)$$, identifying the short-distance
cutoff with $d_0$, since the wall width is also the internal size of the
dipole.
Because the kernel falls as $\rho^{-3}$, the large-$R$ limit is dominated by
nearby points on the ring, $\alpha\to0$ and $\alpha\to2\pi$, where the two
dipoles lie side by side and
$\mathcal V\to[\epsilon(\rho^2+d_0^2)^{3/2}]^{-1}$.
Setting $s=R\alpha$,
\begin{align}
\int d\alpha\,\mathcal V(\alpha)
\simeq
\frac{2}{\epsilon R}
\int_0^\infty ds\,\frac{1}{(s^2+d_0^2)^{3/2}}
=
\frac{2}{\epsilon Rd_0^2}.
\end{align}
Using rotational invariance to write
$E_{\rm dip}=\tfrac12 R^2p_0^2\cdot2\pi\!\int\! d\alpha\,\mathcal V(\alpha)$
and inserting $p_0\simeq eN_wd_0c_G/R$ from \Eq{eq:dipole},
\begin{align}
E_{\rm dip}(R)
\simeq
\frac{2\pi e^2(N_wc_G)^2}{\epsilon R},
\end{align}
in which the wall width cancels between the dipole strength and the
short-distance cutoff.
The overall coefficient is regularization dependent at $O(1)$.
Combining with the surface tension and Zeeman energy gives \Eq{eq:energy}.

\section{Conjugate Aharonov-Casher bands}

In the main text, we used Eq.~\eqref{eq:projector} to construct the charge response arising from a $\bk$-independent mixing of conjugate Chern bands. 
Although analytically tractable and illustrative, the lack of momentum-space winding in the superposition necessarily obstructs spatially local correlations\cite{xie2024theory}, making this particular ansatz energetically unfavorable for any physically relevant short-ranged interaction.
Here, we instead construct a more energetically motivated ansatz by interpolating between the VP and IVC HF reference states as in Eq.~\eqref{eq:interpolation} in the main text.

To implement this program, we need a tractable local model that realizes a pair of time-reversal-related Chern bands while allowing their dispersion and quantum geometry to be tuned. 
We use the adiabatic model introduced to describe the low-lying Chern bands in twisted TMD homobilayers\cite{morales-duranMagicAnglesFractional2024, ShiKhalafMacDonald2024}. 
This model is obtained from the continuum model from a basis transformation that aligns the layer pseudospin with the local moir\'e field, followed by a projection into the diagonal sector.  

Since the moir\'e potential forms a skyrmionic field, the basis change introduces a non-Abelian Berry connection term that generates an effective spatially varying periodic magnetic field, along with a spatially varying residual potential $U(\mathbf{r})$.
The active bands for one of the flavors are thus described by the Hamiltonian for a particle moving in a periodic magnetic field and potential.

\begin{equation}
\hat h_{+}
=
\frac{\Pi^{\dagger}\Pi}{2m}
+U(\br),
\label{eq:ac_hamiltonian}
\end{equation}
where $\mathbf{\Pi} = \hbar \hat{\mathbf{k}}+\mathbf{A}(\mathbf r), \hbar\hat{\mathbf{k}}=-i\nabla, \Pi = \Pi_x + i\Pi_y$, and $\mathbf{A}(\mathbf r)$ generates an emergent, spatially periodic magnetic field
$\mathcal B(\br)=[\nabla\times\mathbf A(\br)]_z$ which carries one flux quantum per unit cell.
$U(\br)$ is a periodic residual scalar potential. 
The other flavor is obtained by time reversal and therefore experiences the opposite emergent magnetic field, producing a pair of conjugate Chern bands.
\begin{figure}[t]
\centering
\includegraphics[width=0.6\columnwidth]{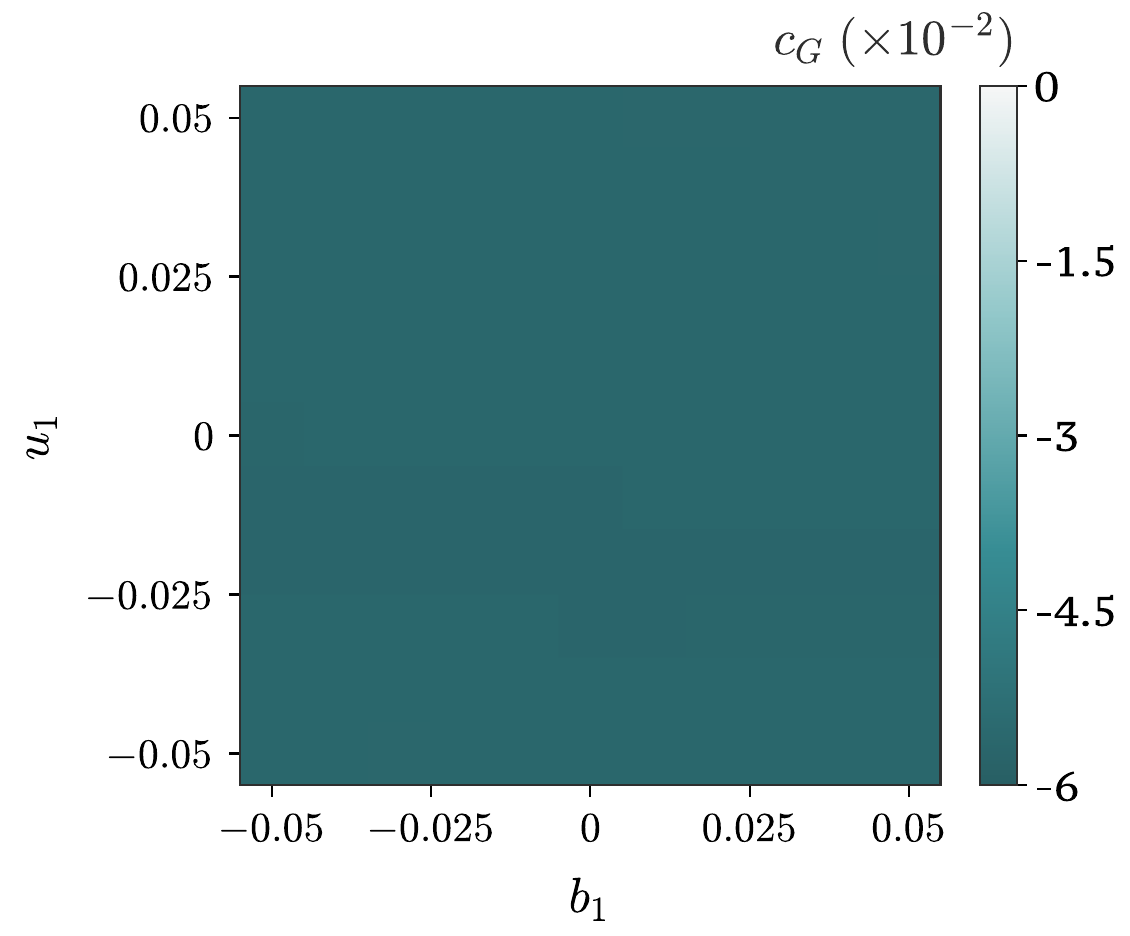}
\caption{
\textbf{$c_G$ calculations for the conjugate AC band setup.}
The figure shows the numerically obtained $c_G$ for the adiabatic model as a function of $u_1,b_1$. We find very little ($<0.5\%$) variation across the phase diagram.}
\label{fig:AC}
\end{figure}

The spatial harmonics of $\mathcal B(\br)$ and $U(\br)$ provide tunable parameters to control the band dispersion and quantum geometry.
In the limit $U(\br)=0$, we obtain the Aharonov-Casher band\cite{morales-duranMagicAnglesFractional2024, ShiKhalafMacDonald2024} which is exactly flat but can have non-uniform quantum geometry.

We rewrite $\mathcal B(\br)=\mathcal B_0+\mathcal B'(\br)$, where
$\mathcal B_0=-2\pi/A_M$ is the average magnetic field, $\omega_c=|\mathcal B_0|/m$, and retain only the first shell harmonics:
\begin{equation}
\begin{aligned}
-\frac{A_M\mathcal B'(\br)}{2\pi}
=
2b_1\sum_{j=1}^{3}\cos(\mathbf G_j\cdot\br),
\\
\frac{U(\br)}{\omega_c}
=
2u_1\sum_{j=1}^{3}\cos(\mathbf G_j\cdot\br),
\label{eq:ac_harmonics}
\end{aligned}
\end{equation}
where $\mathbf G_j$ are three $C_3$-related vectors in the first reciprocal lattice shell ($\mathbf{G}_j=\mathbf{g}_{2j-1}$ defined in Eq.~\eqref{eq:rec_vec}).
The parameter $b_1$ controls the inhomogeneity of the emergent magnetic field
and hence the band quantum geometry. Along the AC line $u_1=0$, this
inhomogeneity deforms the Bloch wavefunctions while preserving an exactly
flat lowest band. A nonzero $u_1$ breaks the AC condition, producing band dispersion
and further modifying the quantum geometry. The point $b_1=u_1=0$ corresponds to the
conjugate lowest Landau level limit.

We diagonalize Eq.~\eqref{eq:ac_hamiltonian} in a basis of
$N_{\rm LL}=8$ magnetic Bloch Landau levels and retain the lowest band from each flavor, on which we project a symmetric dual-gate screened Coulomb interaction:
\begin{equation}
V(q)=
\frac{2\pi V_0 a_M}{q}
\tanh(qd)\,
e^{-(q\ell)^2/2}.
\end{equation}
Here, $V_0$ is the characteristic Coulomb energy, $d$ is the
distance from the layer to each screening gate, and $\ell$ is a short-distance smearing length. 
We use
$V_0/(\hbar\omega_c)=0.3$, $d=30\,{\rm nm}=5.28a_M$, and
$\ell=0.347\,{\rm nm}=0.0611a_M$.

We obtain the VP and IVC reference Hamiltonians from
symmetry-constrained HF calculations as described in the main text, and use Eq.~\eqref{eq:interpolation} to
construct the projectors $P(\bk;\theta,\phi)$ entering the calculation of
$c_G$. 

In order to justify the projection to the lowest AC band, we work in the regime where $u_1, b_1, V_0\ll\omega_c$, where $V_0$ is the characteristic interaction scale.
For Fig.~\ref{fig:AC}, we scan $b_1,u_1\in[-0.05,0.05]$ using a $30\times30$ momentum mesh. 
Throughout this regime, we require that the active bands remain isolated and the interpolating HF Hamiltonian remains
gapped. 
We find that $c_G$ varies by less than $0.5\%$ across the entire range. 
At the conjugate-LL point, $c_G=-0.057$, about $72\%$ of the ideal result $-1/(4\pi)$.

We find issues with HF convergence as well as the appearance of negative indirect gaps for larger ranges of $u_1, b_1$, pointing to the softness of the IVC landscape within the current approximation.

\section{Finite-size scaling of results}
Since our Hartree-Fock calculations are always performed at finite-size, it is important to assess the stability of our result against finite-size effects.
The size-dependence can come from the mesh-size, the cutoff imposed by the number of plane-wave shells used to diagonalize the band-structure, or the number of active bands retained within the Hartree-Fock approximation.

\subsection{Stability to mesh-size}
Unless stated otherwise, any numerical result shown in this paper is for the infinite-size fit extrapolated from the finite-size scaling of the corresponding quantity from the calculations done at mesh-size $N_k=18,19,...,24$.
\begin{figure}[t]
\centering
\includegraphics[width=\columnwidth]{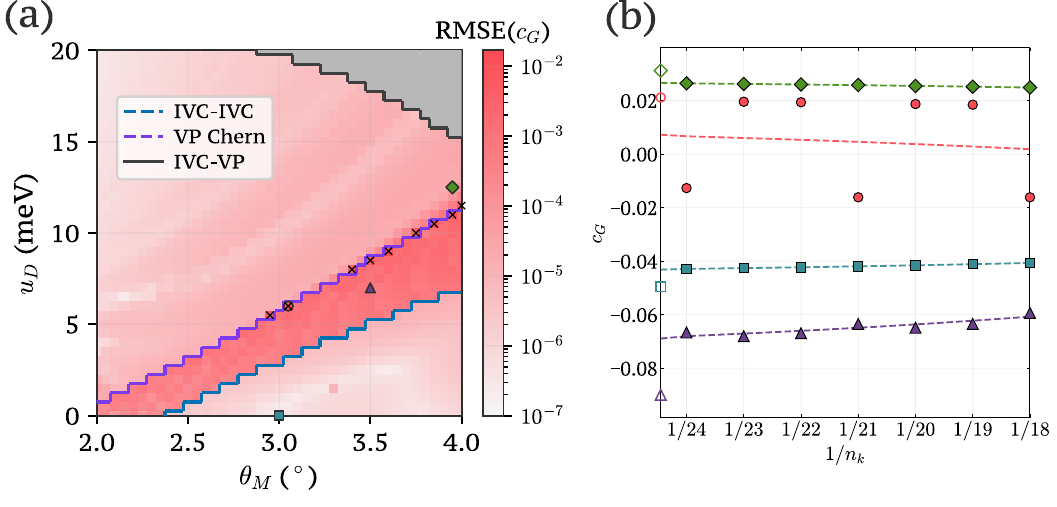}
\caption{
\textbf{Figure showing finite-size scaling and numerical stability analysis.}
(a) shows the RMSE for the linear $1/N_k$ fit for $c_G$ throughout the phase diagram for tMoTe$_2$. (b) selects some representative points from the phase diagram and shows the finite-size scaling fit of $c_G$ values. For the red curve ($\theta_M = 3.05^\circ, u_D = 6\,\rm meV$), the meshes with $c_G<0$ also show $C=1$ for the lowest VP band, whereas the $c_G>0$ meshes show $C=0$ for the lowest VP band.}
\label{fig:mesh_size}
\end{figure}
We find that most of the $c_G$ results behave well under size scaling, and scale linearly with $1/N_k$, with the exception of a few points near the phase boundaries. 
We show the root mean squared error (RMSE) in $c_G$ for the distribution of $c_G$ compared to the finite-size interpolation in Fig.~\ref{fig:mesh_size}.
We show representative points for the different phase regimes in Fig.~\ref{fig:mesh_size}, and find that the numerical instability in $c_G$ is correlated with an instability of the calculated Chern number for the VP state, which shows a mesh-size dependence.
For this point ($\theta_M = 3.05^\circ, u_D = 6\,\mathrm{meV}$), the lowest VP hole band has $C=+1$ when the mesh size is divisible by 3, and correlates with a negative $c_G$, whereas the non-divisible mesh sizes show $C=0$, and have positive $c_G$ values with similar values to other points in the $\rm{VP}, C=0$ part of the phase diagram.
We find a total of 9 such outliers, all lying along the topological phase boundary within the lowest VP band, where the $c_G$ fits show RMSE$>0.005$ stemming from the two different signs of $c_G$ corresponding to the different Chern numbers seen in the lowest VP band.
Thus, these RMSE outliers do not come from random finite-size noise. 
Rather, these are the grid points, where the VP Chern boundary classification is marginal and mesh-dependent, resulting in a highly varying $c_G$.

\begin{figure}[t]
\centering
\includegraphics[width=\columnwidth]{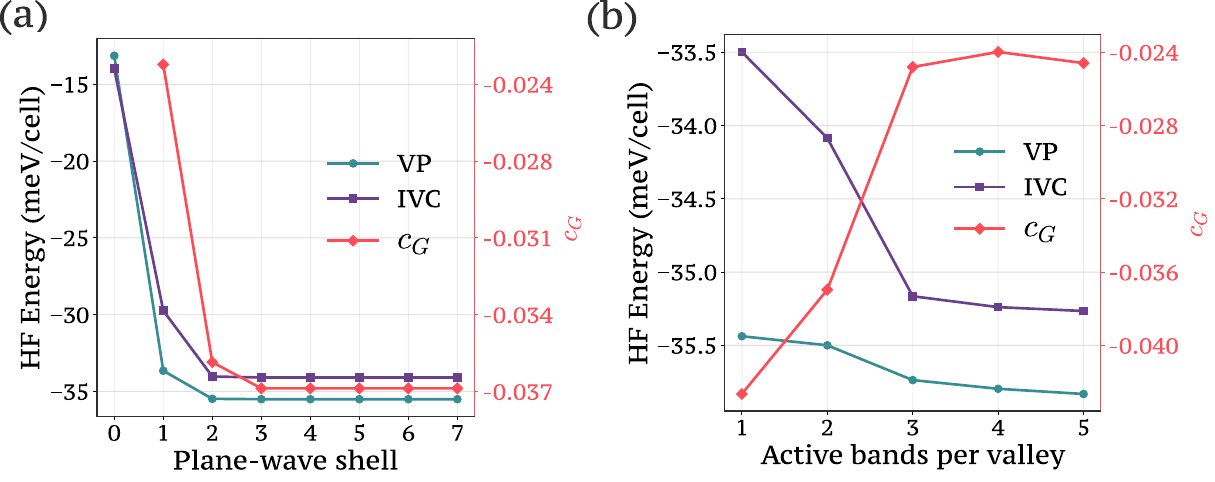}
\caption{
\textbf{Figure showing finite-size scaling and numerical stability analysis.}
We choose a representative point $\theta_M = 3.5^{\circ}, u_D = 0 \,\mathrm meV$, and test the robustness of the results by varying the plane-wave cutoff used to diagonalize the continuum model, and the number of active bands chosen from the continuum model for the projection of interactions. (a) shows the HF energies and $c_G$ for the lowest energy VP and IVC converged solutions as a function of plane-wave cutoff, assuming 2 active bands per valley. (b) shows the variation of the HF energies and $c_G$ as a function of varying the number of active bands per spin-valley.}
\label{fig:cutoff}
\end{figure}
\subsection{Stability to plane-wave and active band cutoff}
We also test the robustness of our results to the various cutoffs assumed in the calculation.
The non-interacting continuum model is diagonalized by choosing plane-wave momenta within a discrete mesh extending up to a few shells. 
By default, we use plane-waves up to the 5\textsuperscript{th} Brillouin zone shell.
As shown in Fig.~\ref{fig:cutoff}, the results are strongly converged by the time we reach the 3\textsuperscript{rd} shell.

The situation is more subtle with the active bands used in the calculation.
By default, we project the interactions onto the low-energy space formed by the 2 lowest energy bands per spin-valley.
In Fig.~\ref{fig:cutoff}, we see that increasing the number of active bands allows states to further lower their energy by hybridizing with higher energy states in the active band subspace.
However, while we see that 2 active bands per spin-valley is not enough to converge the $c_G$, the parameters for the continuum model have been calculated to match the two lowest energy bands per spin-valley obtained from DFT\cite{WangDiverseMagnetic2023}, and including more bands in the projection can also introduce unphysical degrees of freedom to the numerically considered Hilbert space. 
For the purpose of this work, we restrict ourselves to 2 bands per spin-valley, which also drastically reduces the memory and runtime load required for the numerics (memory requirements scale as $N_k^4$).
However, a more microscopically rigorous computation of $c_G$ should compute the continuum model parameters by fitting to the DFT dispersion with more active bands.

\end{document}